\documentclass[aps,preprint]{revtex4}
\usepackage{amsfonts}
\usepackage{amssymb}
\usepackage{amsmath}
\usepackage[font={footnotesize,it}]{caption}
\usepackage{latexsym}
\usepackage{changes}
 \usepackage{color}
\usepackage{hyperref}
\usepackage[capitalise]{cleveref}

\usepackage[caption = false]{subfig}
\usepackage{graphicx}
\usepackage{float}
\usepackage[font=small,labelfont=bf,
   justification=justified,
   format=plain]{caption} 
\hypersetup{
    colorlinks=true,
    linkcolor=red,
    filecolor=magenta, citecolor=red,    urlcolor=blue,
}
\begin{document}

\title{Gravitational lensing in Kerr-Newman anti de Sitter spacetime}
\author{Mert Mangut}
\email{mert.mangut@emu.edu.tr}
\affiliation{Physics Department, Eastern Mediterranean
University, Famagusta, 99628 North Cyprus via Mersin 10, Turkey.}
\author{Huriye G\"{u}rsel}
\email{huriye.gursel@emu.edu.tr}
\affiliation{Physics Department, Eastern Mediterranean
University, Famagusta, 99628 North Cyprus via Mersin 10, Turkey.}
\author{\.{I}zzet Sakall{\i}}
\email{izzet.sakalli@emu.edu.tr}
\affiliation{Physics Department, Eastern Mediterranean
University, Famagusta, 99628 North Cyprus via Mersin 10, Turkey.}

\begin{abstract}
 The method of Rindler and Ishak enables one to study how light is bent in the vicinity of a non-rotating and spherically symmetric gravitational lens. This method mainly aims to investigate the role of cosmological constant in the consequent path. In this paper, we use the extension of Rindler-Ishak method (RIM) in order to evaluate the deflection angle of null geodesics in the equatorial plane of Kerr-Newman anti de Sitter (KNAdS) spacetime. We then use astrophysical data to see the effect of rotation and charge on the bending angle of light for seven distinct stars and two black holes under the assumption of having a KNAdS background with a negative cosmological constant $\Lambda$.

\end{abstract}

\maketitle

\section{Introduction}
In the recent times, gravitational physics and astrophysics have entered a new period: Black holes, which are arguably the most interesting objects of theoretical physics come into reach of direct observation. The phenomenon arising from the deflection of electromagnetic radiation (light) in a field of gravity is called gravitational lensing (GL), and a gravitational lens is an object causing a detectable deflection of light. The main theory of GL, which has been proposed by Liebes and Refsdal \cite{is1,is2}, 
is an astrophysical technique that uses the observed path of light from distant objects to both infer characteristics of foreground mass distributions and image background sources at much higher resolution than is possible with normal telescopes. For further details, the reader is referred to \cite{is3,is4,is5,is6,is7,is8,is8n,is8nn,is8nnn} and references therein.

In fact, GL by black holes began to be observationally important since 1990s. Rauch and Blandford \cite{is9} showed that when a hot spot seen in a black hole’s accretion disk or jet passes through caustics of the Earth’s past light cone (caustics produced by the spacetime curvature of the hole), the brightness of the X-rays of hot spots will be subject to sharp oscillations with informative shapes. This has motivated a number of quantitative studies of the Kerr metric’s caustics \cite{is10,is11}.

One of the main motivations behind a great many of theoretical cosmological studies is to establish a model that fits the astronomical data. The potential candidates of such a cosmological model are aimed to be used for
explaining the formation and evolution of the Universe. In 1998 \cite{is12}, the observational evidence of a high red-shift type Ia supernova suggested that the expansion of our cosmos is actually accelerating in nature \cite{is13}, which convinced researchers to rehabilitate the cosmological constant ($\Lambda$) concept originally proposed by Einstein. The effects of the variation of $\Lambda$, in the astrophysical scenario, have been studied in great detail by \cite{is14n,is15n}. While many plausible models have been proposed with the intention of explaining this accelerated expansion, the cold dark matter model with $\Lambda$ ($\Lambda$CDM model) seems to be highly consistent with the observational data \cite{is14}. For this reason, Einstein's equations need to be modified with $\Lambda$ at the cosmological scale.

To fully grasp the dynamical behavior of the Universe, it would be a good starting point to treat classical observations such as the accelerated expansion of the Universe as the low energy realizations of an underlying quantum mechanical mechanism satisfying the constraints of general relativity. As previously stated, the $\Lambda$CDM model already provides us with satisfying answers once large scales are of concern. However, the current observations suggest that there exist convincing reasons to start looking for alternatives or modified versions of this model. Treating the $\Lambda$CDM model as the cosmological framework of the Universe leads to problems such as the $\Lambda$-fine tuning, the Hubble tension and the cosmic coincidence \cite{new3}. One of the probable reasons for these inconsistencies may be the assumption of having a positive $\Lambda$. In \cite{new2}, Hartle, Hawking and Hertog stated that quantum cosmology allows a non-trivial connection between the observed cosmological parameters and those of the underlying theory, since both the backgrounds and the fluctuations are treated quantum-mechanically. They have also added that fundamental theories possessing negative $\Lambda$ can be consistent with our low-energy observations of a classical, accelerating universe (one may also find \cite{Hawking2} interesting). Furthermore, in \cite{new1}, Visinelli et al. proposed a cosmological model, namely cCDM, with negative $\Lambda$. Although their results show that $\Lambda$CDM is favoured over string-inspired cCDM, the authors indicate that the cCDM model is also consistent with the observational data. In 2021, their study was extended by Sen, Adil and Sen \cite{Sen} with the addition of cosmic microwave background and lensing data into the picture. Within this extended model, Sen et al. have shown that taking $\Lambda$ as negative is always preferred over the $\Lambda$CDM model. Over the last few years, models or implications of negative $\Lambda$ has been quite popular especially due to AdS/CFT correspondence (see, for instance, \cite{ads,is15,is16}).  In addition to the negative $\Lambda$ models, there exists an intriguing study in which a sign-switching $\Lambda$ is proposed. The authors refer to their model as the $\Lambda_{s}$CDM model.

In this work, we have chosen $\Lambda$ to be negative so as to allow string-inspired applications and check the effect of this choice on the deflection angle of light. Studies on the effect of $\Lambda$ on local phenomena such
as null geodesics, time delay of light, gravitational time advancement, and the perihelion precession, have
attracted attention over the last decades. In addition to all these, the role of $\Lambda$ in GL is also considered as an interesting topic to dive into. In \cite{isIslam},  Islam claimed the geodesic equation of light in a spherically symmetric spacetime would not contain any $\Lambda$ term, which would in turn imply $\Lambda$ does not alter the bending of light. However, Rindler and Ishak \cite{is25} came up with a new method, which will be referred to as RIM from this point onwards, to see if this was really the case. In contrary to the previous studies, they showed that $\Lambda$ does affect the path that light follows. Furthermore, in \cite{26}, Ishak has derived the effect of $\Lambda$ on the bending angle from both the integration of the gravitational potential and Fermat’s principle. In addition to Rindler and Ishak, other researchers have also conducted important studies on GL which are related to rotating and twisting universes, nonlinear electrodynamics, galactic halo, monopoles etc. (one may refer to \cite{is26,izz,is27,is28,is29,is30,is31,is32,ariel,ellis,new7,new8} ).

Today, RIM is considered as a rather convenient methodology (see, for instance, \cite{isHeydari}), and based on this fact, we will be using the extended version of this method to investigate the GL phenomenon in the stationary KNAdS geometry, which can be used for modeling stars and black holes [see Secs. \eqref{sect4} and \eqref{sect5}]. Nonetheless, there is a conceptual issue with the RI approach: it should be noted that using an observer with fixed Boyer-Lindquist coordinates leads to different contributions of $\Lambda$ to the light deflection angle than considering a perspective from cosmological co-moving coordinates. Thus, one shall pay attention to projective equivalence and record that we will be continuing our analysis in Boyer-Lindquist coordinates. For further information in this manner, the reader is referred to \cite{isPark}.

It is also worth noting that GL can be treated as a tool to attempt further understanding the properties of certain astrophysical objects.  Recently, astrophysicists have been focusing on perfecting the mm-intermerometer imaging of the black hole shadows and discs, especially the mm-intermerometer imaging of the black hole ($Sgr$ $A^{*}$) in the middle of Milky Way Galaxy \cite{is21,is22}. These studies result in a mm-interferometric system, which is the so-called "Event Horizon Telescope (EHT)" \cite{is23}. In 2019, the EHT collaboration delivered the first image of $M87$ black hole -- the supermassive object in the center of the $M87$ galaxy \cite{is24}. However, the gravitational lensing of a rotating compact object in AdS spacetime has not been yet fully understood. In astrophysics, the celestial compact objects such as quark stars and neutron stars are treated as test beds for analyzing various properties of gravity. In this study, we consider a class of compact stars, namely $PSR J 1748-2446ad,$ $PSR B 1937+21$, $PSR J 1909-3744$, $PSR J 0737-3039 A$ and $PSR B 1534+12$ with radii $20.1$ $km$, $20.2$ $km$, $31.1$ $km$, $133.6$ $km$ and $165.9$ $km$ , respectively. In addition to black holes, we picked neutron stars as the lenses causing GL, since neutron stars can in some cases be more compact than certain black holes and it is of great interest to check the bending effects of these astronomical objects \cite{son}.

There exist a large body of literature on null geodesics and GL by generalizations of the Kerr spacetime, including full analytical solutions for Kerr AdS \cite{isHackmann}  and some generalizations to the Plebanski-Demianski class which contains KNAdS as a special case. However, the main aim of the present paper is to use the extended version of RIM in KNAdS background and inspect the effect of rotation and charge on the bending angle of light, once it passes nearby a set of astronomical objects of our choice. Also, for using general relativistic aberration formula in the Kerr-de Sitter spacetime, one can check \cite{new5}; whereas the exact solutions for the deflection angle of
equatorial light rays (in the conventional approach for the computation of the deflection angle) in the Kerr-Newman and Kerr-Newman-(anti) de Sitter black hole spacetimes were obtained, in terms of multivariable Appell-Lauricella hypergeometric functions and elliptic Weierstraß functions in \cite{new6}. At this stage, one may question why we consider the KNAdS solution instead of, for instance,  Hartle-Thorne \cite{Hartle:1968si} or Manko \cite{Manko:2000ud,Manko:2000sg} spacetimes. In this regard, we should point out that there are compelling evidences indicating the outer regions of rotating stars can plausibly be considered as Kerr-like spacetimes \cite{Berti:2004ny,Nathanail:2017wly}. Furthermore, it can be deduced from Ref. \cite{Korea} that the Hartle-Thorne metric reduces to the Kerr-like spacetime once the physical parameters are chosen as $\theta=\frac{\pi}{2}$ and $\frac{a^2}{2r^2}\ll 1$; $r\rightarrow r\left[1-\frac{a^2}{2r^2}\left \{\left(1+\frac{2M}{r} \right)\left(1-\frac{M}{r} \right) \right\} \right]\approx r $ and $\theta \rightarrow \theta=\pi/2$, which coincide with our choices. (Note that the physical values to be used in our study are taken from \cite{Korea}; see also Table \ref{table:nonlin1}). The latter remark can be best seen from the fifth equation of \cite{Korea}. In addition, the  Kerr spacetime geometry, which will be the same metric (for $\Lambda\rightarrow0$ amd $e\rightarrow 0$ ) to be employed in this study, can also be used for analyzing the gravitational redshift effect of the same rotating stars \cite{Dubey:2014gia}. Based on these findings, the charged stars and black holes of this study will be described by the KNAdS metric. Besides, we will employ the KNAdS metric to account for the cosmological constant in our calculations.

The brief outline of this paper is the following. Section
\ref{sect2} introduces the KNAdS metric and represents some of its physical features. In Sec. \ref{sect3}, we describe a generic formalism for computing the bending angle of light for 4-dimensional rotating geometries. Section \ref{sect4} is devoted to the computations of the GL in the KNAdS spacetime. In Sec. \ref{sect5}, we use the obtained theoretical results for the astrophysical objects. Conclusions are presented in Sec. \ref{sect6}.

\section{KNA$d$S SPACETIME IN FOUR DIMENSIONS} \label{sect2}
The exact solution of the Einstein-Maxwell field equations with $\Lambda$, which
describes stationary spacetimes in 4-dimensional spacetime with asymptotic dS and/or AdS behavior, was found long ago by Carter \cite{is33} . The corresponding metric in the Boyer-Lindquist
coordinates is given by \cite{Zhang:2018ocv}
\begin{equation}
d s^{2}=-\frac{\Delta_{r}}{\rho^{2}}\left(d t-\frac{a \sin ^{2} \theta}{\Xi} d \varphi\right)^{2}+\frac{\Delta_{\theta} \sin ^{2} \theta}{\rho^{2}}\left(a d t-\frac{r^{2}+a^{2}}{\Xi} d \varphi\right)^{2}+\rho^{2}\left(\frac{d r^{2}}{\Delta_{r}}+\frac{d \theta^{2}}{\Delta_{\theta}}\right), \label{metric}
\end{equation}
where
\begin{equation}
\begin{split}
\varrho^{2}&=r^{2}+a^{2} \cos ^{2} \theta, \quad \Delta_{\theta}=1-\frac{a^2}{l^2} \cos ^{2} \theta, \quad  \Xi=1-\frac{a^2}{l^2}, \\ \Delta_{r}&=\left(r^{2}+a^{2}\right)\left(1+\frac{r^2}{l^2} \right)-2 m r+e^{2},    
\end{split}
\end{equation}
in which $a$ denotes the rotational parameter, $e$ and $m$ represent the charge and mass parameters, respectively, and $l$ is \textit{the curvature radius determined via} $\Lambda=-3 l^{-2}$ \cite{Zhang:2018ocv,is17}.
The mass $M$ and the angular momentum $L$ of the KNAdS can be obtained with the aid of Komar integrals by using the Killing vectors $\partial_{t} / \Xi$ and $\partial_{\phi}{ }$ \cite{is35}. Taking AdS space as reference background, one can get
\begin{equation}
M=\frac{m}{\Xi^{2}}, \quad L=\frac{a m}{\Xi^{2}}. 
\end{equation}
Similarly, the charge $Q_e$ of the KNAdS can be obtained by computing the flux of the electromagnetic field tensor at infinity \cite{is36}:
\begin{equation}
Q_{e}=\frac{e}{\Xi}.
\end{equation}.

Metric (\ref{metric}) is the solution to the Einstein-Maxwell field equations with the following electromagnetic vector potential (1-form):
\begin{equation}
A_{t}=-\frac{e r}{\rho^{2}}, \quad A_{\varphi}=\frac{a  e \sin ^{2} \theta}{\rho^{2} \Xi}, \label{em1}
\end{equation}
in which the angular (magnetic) component $A_{\varphi}$ is due to the the rotation of the black hole. On the other hand, one can check that the electromagnetic potential yields the following field strength tensor:
Meanwhile, the vierbein fields \cite{is17} of the metric \eqref{metric} can be defined as follows
\begin{equation}
\begin{split}
\mathrm{e}^{0}&=\frac{\sqrt{\Delta_{r}}}{\rho}\left(\mathrm{d} t-\frac{a \sin ^{2} \theta}{\Xi} \mathrm{d} \varphi\right), \quad \mathrm{e}^{1}=\frac{\rho}{\sqrt{\Delta_{r}}} \mathrm{~d} r, \\
\mathrm{e}^{2}&=\frac{\rho}{\sqrt{\Delta_{\theta}}} \mathrm{d} \theta, \quad \mathrm{e}^{3}=\frac{\sqrt{\Delta_{\theta}} \sin \theta}{\rho}\left(a \mathrm{~d} t-\frac{r^{2}+a^{2}}{\Xi} \mathrm{d} \varphi\right).
\end{split}
\end{equation}
On the other hand, one can check that the electromagnetic potential \eqref{em1} yields the following field strength tensor:
\begin{equation}
F=\frac{e\left(\rho^2-2 r^{2}\right)}{\rho^{4}}\left(d t-\frac{a \sin ^{2} \theta}{\Xi} d \varphi\right) \wedge d r+\frac{e r a \sin 2 \theta}{\rho^{4}}\left(a d t-\frac{r^{2}+a^{2}}{\Xi} d \varphi\right) \wedge d \theta,
\end{equation}
or
\begin{equation}
F=\frac{e\left(\rho^{2}-2r^{2}\right)}{\rho^{4}} (\mathrm{e}^{0} \wedge \mathrm{e}^{1}) -\frac{2 e r a \cos \theta}{\rho^{4}}(\mathrm{e}^{2} \wedge \mathrm{e}^{3}).
\end{equation}
One can also compute the non-zero  contravariant components
of the electromagnetic field tensor to find, in particular, the magnetic fields produced by the KNAdS spacetime:
\begin{equation}
\begin{array}{ll}
F^{tr}=\frac{e\left(r^{2}+a^{2}\right)}{\rho^{6}}\left(2 r^{2}-\rho^2\right), & F^{t\theta}=-\frac{e a^{2} r }{\rho^{6}}\sin 2 \theta, \\ \\
F^{r\varphi}=\frac{e a\left(2 r^{2}-\rho^2\right)}{\rho^{6}} \Xi, & F^{\theta\varphi}=\frac{2 e a r}{\rho^{6}} \Xi \cot \theta.
\end{array} \label{Bfield}
\end{equation}
The magnetic field components of the KNAdS can now be read from Eq. \eqref{Bfield} by using the following expression \cite{EllisBook1,EllisBook2}:
\begin{equation}
B_{\mu}=\frac{1}{2}E_{\mu \nu \alpha \tau} u^{\nu} F^{\alpha \tau}
\end{equation}
where $u^{\nu}$ denotes the 4-velocity vector and the covariant Levi-Civita tensor (also known as the Riemannian volume form) is represented by $E_{\mu \nu \alpha \tau} \equiv|g|^{1 / 2} \varepsilon_{\mu \nu \alpha \tau}$ with $\varepsilon_{tr\theta\phi}=+1$ \cite{clem1,clem2}. Therefore, the non-zero components of the magnetic fields can be obtained as follows:
\begin{equation}
B_{r}=\frac{2 e a r\left(r^{2}+a^{2}\right) \cos \theta }{\rho^{4}\Delta_{r}}. \label{11}
\end{equation}
\begin{equation}
B_{\theta}=\frac{eal^{2}\left(\rho^2-2r^{2}\right) \sin \theta}{\rho^{4}(a^2\cos^2\theta -l^2)}, \label{12}
\end{equation}

which are the components of the magnetic field's magnitude expression: $B=\sqrt{B_{r}^2+B_{\theta}^2}$. It is also worth noting that the expressions for the magnetic and electric field components for the Kerr-Newman spacetime in the presence of the cosmological constant for a ZAMO (zero-angular-momentum observer) were derived in \cite{new6}. In the limit of very small $\Lambda$, when the detector is sufficiently far away from the black hole such that $a \ll r$, Eq. \eqref{11} approximates to
\begin{equation}
B_{r}\approx\frac{2ea\cos\theta}{r^{3}}+\frac{4eam\cos \theta}{r^{4}}+\mathcal{O}\left(\frac{1}{r^{5}}\right), \label{moment}
\end{equation}

which trivially returns to the radial component of the magnetic field for the Kerr-Newmann solution \cite{grbook}. 

It is also worth noting that the Biot-Savart law of electrodynamics implies a charge $e$ with mass $m$ on a circular orbit with angular momentum $\vec{L}$ has the following magnetic dipole moment:
\begin{equation}
\vec{\mu}=\mathcal{X} \frac{e \vec{L}}{2m},
\end{equation}
where $\mathcal{X}$ is the gyromagnetic moment \cite{grbook}. The magnetic dipole moment $\vec{\mu}$ creates a dipole field that can be expressed as follows:
\begin{equation}
\vec{B}=\frac{3\left(\vec{\mu} \cdot \vec{e}_{r}\right) \vec{e}_{r}-\vec{\mu}}{r^{3}}, \label{xx}
\end{equation}
In turn, the radial component of this field is given by $B_{r}=\vec{B} \cdot \vec{e}_{r}=2 \vec{\mu} \cdot \vec{e}_{r} / r^{3}$. A comparison of the radial magnetic field (considering the leading order term) from Eq. \eqref{moment} with expression \eqref{xx} yields
\begin{equation}
\vec{\mu}=e \vec{a}=\frac{e \vec{L}}{m}=2 \frac{e \vec{L}}{2m},
\end{equation}
showing that a charged slow-rotating KNAdS spacetime (with $\Lambda\ll$) has a gyromagnetic moment of $\mathcal{X}=2$.

The horizons of metric \eqref{metric} are found by the (positive) roots of the equation $\Delta_{r}=0$ \cite{is17}:
\begin{equation}
\begin{split}
\Delta_{r} &=\left(r^{2}+a^{2}\right)\left(1-\frac{1}{3} \Lambda r^{2}\right)-2 m r+e^{2} \\
&=-\frac{1}{3} \Lambda\left[r^{4}-\left(\frac{3}{\Lambda}-a^{2}\right) r^{2}+\frac{6 M}{\Lambda} r-\frac{3}{\Lambda}\left(a^{2}+e^{2}\right)\right] \\
&=-\frac{1}{3} \Lambda\left(r-r_{++}\right)\left(r-r_{--}\right)\left(r-r_{+}\right)\left(r-r_{-}\right)=0,
\end{split}
\end{equation}
 where $r_{++}$ and $r_{--}$ are a pair of complex conjugate roots, $r_{+}$ and $r_{-}$ are two real positive roots with $r_{+}>r_{-}$. Thus, $r=r_{+}$ is the event horizon. 
 
 The Hawking temperature of the KNAdS black hole is given by \cite{is34}
\begin{equation}
T_{H}=-\frac{\Lambda}{12 \pi\left(r_{+}^{2}+a^{2}\right)}\left(r_{+}-r_{++}\right)\left(r_{+}-r_{--}\right)\left(r_{+}-r_{-}\right). \label{TH}
\end{equation}
Since $\Lambda<0$, $r_{+}$ and $r_{-}$ have positive values. Furthermore, considering that
$r_{+}>r_{-}$ and $r_{++}$ and $r_{--}$ are complex conjugates, one can infer that $T_{H}$ has a positive value, as expected. In fact, Eq. (\ref{TH}) can be rewritten as \cite{is35}
\begin{equation}
T_{H}=\frac{3 r_{+}^{4}+\left(a^{2}+\ell^{2}\right) r_{+}^{2}-\ell^{2}\left(a^{2}+e^{2}\right)}{4 \pi \ell^{2} r_{+}\left(r_{+}^{2}+a^{2}\right)}.
\end{equation}

\section{Review of Generalized Rindler and Ishak Formalism for Finding Bending of Light in Rotating Spacetimes } \label{sect3}

\begin{figure}[htp]
  \includegraphics[width=0.8\textwidth]{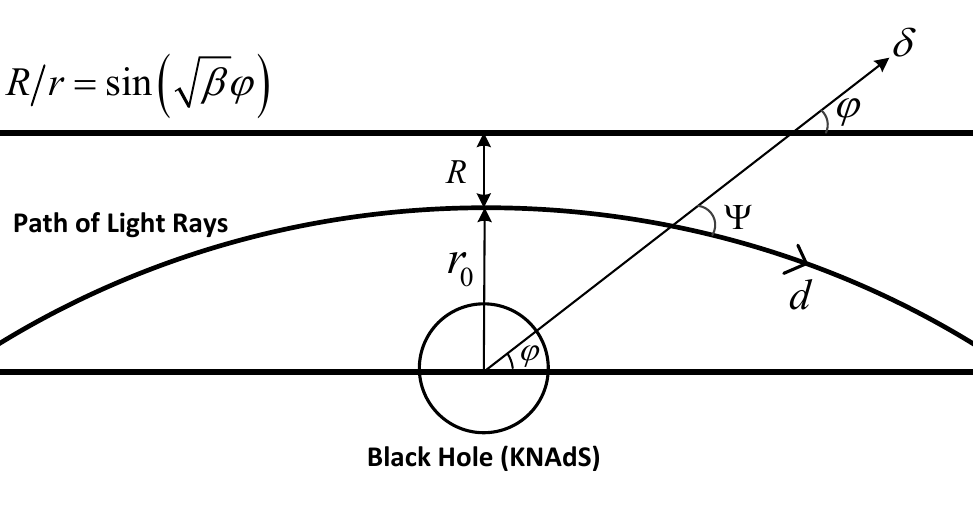}

  \captionof{figure}{In this figure, the gravitational lensing effect created by a massive object in space is shown with the relevant parameters.}  \label{fig1}
  \end{figure}

In 2007, Rindler and Ishak (RI) investigated gravitational lensing within Schwarzschild Sitter (SdS) geometry so as to see whether $\Lambda$ plays any role in the overall bending of light \cite{is25}. Their method is now very frequently used in analyzing gravitational lensing for geometries with non-asymptotic flatness. Moreover, this method is further generalized in \cite{is29} for rotating space-times. Let us now give a brief summary of the generalization made for geometries with rotation. In general, the geometry of rotating spacetimes can be represented with the line element
\begin{equation}
ds^{2}=f(r)dt^{2}+2g(r)dtd\varphi-h(r)dr^{2}-p(r)d\varphi^{2}. \label{20}
\end{equation}   
Note that the metric is written for constant $\theta=\pi/2$. 
The RIM involves generalizing the inner product to curved spaces, where the invariant angle between two vectors is found. In the light of this information, the angle between two coordinate directions $d$ and $\delta $, as shown in Fig. \ref{fig1} is given by 

\begin{equation}
\cos \left( \psi \right) =\frac{d^{i}\delta _{i}}{\sqrt{\left(
d^{i}d_{i}\right) \left( \delta ^{j}\delta _{j}\right) }}=\frac{%
g_{ij}d^{i}\delta ^{j}}{\sqrt{\left( g_{ij}d^{i}d^{j}\right) \left(
g_{kl}\delta ^{k}\delta ^{l}\right) }}, \label{21}
\end{equation}

where $g_{ij}$ 	illustrates the concerned metric tensor. For the analysis, the path of light at the equatorial plane ($\theta =\pi /2$ ) is described in two-dimensional curved $(r,\varphi )$ space. We have chosen equilateral plane so as to keep the direction of angular momentum conserved \cite{Sean}. Hence, when a constant time interval is concerned, we can write
\begin{equation}
dl^{2}=h(r)dr^{2}+p(r)d\varphi ^{2}. \label{22}
\end{equation}%
The general null geodesics equation can be written as
\begin{equation}
\left( \frac{dr}{d\varphi }\right) ^{2}=\frac{K(r)}{h(r)\left[g(r)E+f(r)L)\right]^{2}}\left[p(r)E^{2}-f(r)L^{2}-2g(r)LE\right]. \label{23}
\end{equation}
where $E$ and $L$ stand for energy and angular momentum constants, respectively.
The relevant constants of motion are%
\begin{equation}
\frac{dt}{d\lambda }=\left[ \frac{-p(r)}{K(r)} \right]E+ \left[ \frac{g(r)}{K(r)} \right] L,\text{ \ \ \
\ \ }\frac{d\varphi }{d\lambda }= \frac{-g(r)E-f(r)L}{K(r)} \label{24}
\end{equation}%
with $\lambda $ representing the affine parameter and $K(r)=g^{2}(r)+p(r)f(r)$.  If we apply standard change of variable process with $u=\frac{1}{r}$, \ Eq. \eqref{23} reduces to
\begin{equation}
\frac{d^{2}u}{d\varphi ^{2}}=2u^{3}\kappa(u)+\frac{u^{4}}{2}\frac{d\kappa(u)%
}{du}, \label{25}
\end{equation}%
where $\kappa(u)=\frac{K(r)}{h(r)\left[g(r)E+f(r)L)\right]^{2}}\left[p(r)E^{2}-f(r)L^{2}-2g(r)LE\right]$. 
Assigning symbol $d$ for the direction of the path taken by light and keeping the coordinate line $\varphi =$ constant $\delta ,$ we can state
\begin{eqnarray}
d &=&\left( dr,d\varphi \right) =\left( A,1\right) d\varphi \text{ \ \ \ \ \
\ \ }d\varphi <0,  \notag \\
\delta &=&\left( \delta r,0\right) =\left( 1,0\right) \delta r,
\end{eqnarray} \label{26}
in which
\begin{equation}
A(r,\varphi )\equiv \frac{dr}{d\varphi }. \label{27}
\end{equation}
Substituting these constraints back into Eq. \eqref{21} leads to

\begin{equation}
\tan \left( \Psi \right) =\frac{\left[ h^{-1}(r)p(r)\right] ^{1/2}}{%
\left\vert A(r,\varphi )\right\vert },  \label{28}
\end{equation}%
 Furthermore, the one-sided bending angle is given by $\epsilon =\Psi
-\varphi .$

\section{Gravitational Lensing in KNA$d$S Spacetime} \label{sect4}

To derive the associated light ray equation, one can have a closer look at Eq. \eqref{25}. Keeping in mind that the signature of metric \eqref{metric} is $+2$, the light ray equation can then be written as
\begin{equation}
\frac{d^{2}u}{d\varphi ^{2}}+\beta u=\frac{3Mu^{2}}{\alpha^{2}}-\frac{2e^{2}u^{3}}{\alpha^{2}},  \label{29}
\end{equation}

where the dimensionless parameters are $\beta=-\frac{\Lambda a^{2}}{3 \alpha^{2}}+\frac{L+Ea}{\alpha^{2} \left(L-Ea \right)}$ and   $\alpha=1+\frac{1}{3}\Lambda a^{2}$.

By taking advantage of the linear solution of Eq.(29) $u=\frac{sin(\sqrt{\beta}\varphi)}{R}$ with $R$ representing the impact parameter, an approximate solution is achieved. Once substituted in Eq. \eqref{29}, the approximate solution of the differential equation is found as  
\begin{equation}
\begin{aligned}
u(\varphi)=\frac{\sin(\sqrt{\beta}\varphi)}{R}+&\frac{1}{4R^{3}}\left\{\left((1+\cos^{2}(\sqrt{\beta}\varphi)\right)4MR+e^{2}\left(   \frac{3\varphi \cos(\sqrt{\beta}\varphi)}{\sqrt{\beta}\alpha^{2}}-\frac{ sin(\sqrt{\beta}\varphi)\cos^{2}(\sqrt{\beta}\varphi)}{\beta\alpha^{2}}\right. \right.\\
&\left. \left. -\frac{2\sin(\sqrt{\beta}\varphi)}{\beta\alpha^{2}}\right)\right\}.
\end{aligned} \label{30}
\end{equation}
Consequently,  Eq. \eqref{27} becomes
\begin{equation}
 \begin{aligned}
A(r,\varphi)=&\frac{r^{2}}{4R^{3}} \left\{   4MR\sqrt{\beta}\sin(2\sqrt{\beta}\varphi) +e^{2} \left(\frac{\cos^{3}(\sqrt{\beta}\varphi)}{\sqrt{\beta}\alpha^{2}} -\frac{sin(2\sqrt{\beta}\varphi)sin(\sqrt{\beta}\varphi)  }{\sqrt{\beta}\alpha^{2}}\right. \right.\\
&\left. \left.     \frac{3\varphi \sin(\sqrt{\beta}\varphi)}{\alpha^{2}}-\frac{cos(\sqrt{\beta}\varphi)}{\sqrt{\beta}\alpha^{2}}                                     \right)  \right \}-\frac{r^{2}}{R}\sqrt{\beta}cos(\sqrt{\beta}\varphi).
\end{aligned} \label{31}
\end{equation}

To check the effect of the geometry and physical parameters (charge, mass and spin) of the concerned black hole model on the closest approach distance equation, $r_{0}$\ is analyzed at $\varphi =\pi /2$. 
The reciprocal of the closest approach distance is, in turn, evaluated to be

\begin{equation}
\begin{aligned}
\frac{1}{r_{0}}=&\frac{sin(\sqrt{\beta}\pi /2)}{R}+\frac{1}{4R^{3}}\left\{\left((1+cos^{2}(\sqrt{\beta}\pi /2)\right)4MR+ \right.\\& \left. e^{2}\left(   \frac{3\pi cos(\sqrt{\beta}\pi /2)}{2\sqrt{\beta}\alpha^{2}}-\frac{ sin(\sqrt{\beta}\pi /2)cos^{2}(\sqrt{\beta}\pi /2)}{\beta\alpha^{2}}-\frac{2sin(\sqrt{\beta}\pi /2)}{\beta\alpha^{2}}\right)\right\}.
\end{aligned}   \label{32}
\end{equation}

In order for being able to check astrophysical applications of our results, let us investigate what happens for $\varphi =0$. For the cases when $\frac{M}{R}<<1$ and $\Lambda R^{2}<<1$  \cite{25,27,is25},
the radial coordinate and its first derivative with respect to $\varphi$ can be written as
\begin{equation}
r\approx \frac{\beta\alpha^{2}R^{2}}{2M},\text{ \ \ \ \ \ } A(r,\varphi=0)%
\approx- r^{2}\frac{\sqrt{\beta}}{R}.    \label{33}
\end{equation}
Then, we obtain
\begin{equation}
 \begin{aligned}
\tan\epsilon =\tan\Psi _{0}\simeq \frac{2M}{\beta^{3/2}\alpha^{3}R}&\left\{ 1-\frac{\beta^{2}\alpha^{5}R^{4}\Lambda}{%
12M^{2}}-\frac{ 4M^{2}}{\alpha^{2}\beta R^{2}}-\frac{a^{2}\Lambda}{3}\left[ 1+\frac{2}{3\alpha^{2}}    \right ]  \right.\\
 &\left. +\frac{4M^{2}}{\beta^{2}\alpha^{4}R^{4}}\left[ a^{2}+e^{2}+\frac{2a^{2}}{\alpha^{2}}-   \frac{2a^{4}\Lambda}{3\alpha^{2}}- \frac{2a^{4}\Lambda}{3\beta^{2}}   \right]    \right\}^{1/2}.
 \end{aligned} \label{34}
\end{equation}

If we use the standard expansion of square root, the one-sided bending angle can be written as

\begin{equation}
 \begin{aligned}
\tan\epsilon =\tan\Psi _{0}&\simeq \frac{2M}{\beta^{3/2}\alpha^{3}R}\left\{ 1-\frac{\beta^{2}\alpha^{5}R^{4}\Lambda}{%
24M^{2}}-\frac{ 2M^{2}}{\alpha^{2}\beta R^{2}}-\frac{a^{2}\Lambda}{6}\left[ 1+\frac{2}{3\alpha^{2}}    \right ]  \right.\\
 &\left. +\frac{2M^{2}}{\beta^{2}\alpha^{4}R^{4}}\left[ a^{2}+e^{2}+\frac{2a^{2}}{\alpha^{2}}-   \frac{2a^{4}\Lambda}{3\alpha^{2}}- \frac{2a^{4}\Lambda}{3\beta^{2}}   \right]    \right\} +\mathcal{O}\left( \frac{M^{9}a ^{4}}{\alpha^{19}\beta^{19/2}R^{17}}\right)   .
 \end{aligned} \label{35}
\end{equation}

Note that when $a\rightarrow{0}$, Eq.\eqref{35} reduces to the one-sided bending angle for the linear case in \cite{is26}. 

At this point, one needs to determine how to deal with $\beta$. As can be seen from Eq. \eqref{29}, this dimensionless parameter contains information regarding the light ray propagating through the spacetime of our concern. If we use the definition of impact parameter for rewriting of this dimensionless parameter, considering $R\equiv L/E$ \cite{izz}, we find
\begin{equation}
    \beta=-\frac{\Lambda a^{2}}{3 \alpha^{2}}+\frac{R+a}{R-a}. \label{36}
\end{equation}

\section{Applications in Astrophysics } \label{sect5}

This section is reserved for astrophysical applications of our evaluations. We have used observational data to check the contribution of our answers to the bending angles due to seven observed
charged compact stars ($PSR J 1748-2446ad,$ $PSR B 1937+21$, $PSR J 1909-3744$, $PSR J 0737-3039 A$ and $PSR B 1534+12$), and two black holes: Sgr $A^{*}$ and M87. The properties of the relevant compact stars are tabulated in Table I  \cite{17,18,new19,new20,new21,Dubey:2014gia}.

\begin{table}[th]
\caption{The numerical values of the masses $(M)$, radii $(R_{Star})$, rotation parameter $(a)$ , angular velocity $(\Omega)$, period $(P)$ and period derivative $(\dot{P})$ of the
compact stars. Here $M_{\odot }$ denotes the mass of the sun.} 
\label{table:nonlin1}
\centering
\begin{tabular}{|c|c|c|c|c|c|c|c|c|}
\hline
Compact Stars & $M/M_{\odot }$ & $R_{Star}$ (km) & a (km)&$\Omega$ $\times 10^{3}$ (rad/s)&$P$ $\times 10^{-4}$(s)&$\dot{P}$ $\times 10^{-18}$ \\ \hline\hline
PSR J 1748-244ad  & $1.350$ & $20.1$ & $2.423$ &$4.498$ & $13$ & $0.60$\\ \hline
PSR B 1937+21  & $1.350$ & $20.2$ & $%
2.194$ &$4.033$ & $15$ & $%
0.16$\\ \hline
PSR J 1909-3744  & $1.438$ & $31.1$ & $2.746$ & $2.130$ & $29$ & $0.01$\\ \hline
PSR J 0737-3039 A  & $1.340$ & $133.6$ & $6.582$ & $0.276$ & $22$ & $1.74$\\ \hline
PSR B 1913+12  & $1.340$ & $165.9$ & $6.080$& $0.165$ & $379$ & $2.42$ \\ \hline
\end{tabular}
\end{table}

 The numerical values of the mass, charge and radius of the
 black holes $Sgr$ $A^{*}$ and $M87$  are taken from \cite{19,new9,21,22} and $\Lambda$ will be taken as $\Lambda=-1.1\times10^{-52}$ $m^{-2}$ \cite{Planck}. According to studies \cite{19,new9},
$Sgr$ $A^{*}$ having a mass of $4.1\times 10^{6} M_{\odot }$ with the Schwarzschild radius $ 1.27 \times 10^{10} m $ holds an overall charge of  the order $10^{15} C$. We have picked these values for our analysis. On the other hand, for M87; the observational data used are $M_{M 87}=6.5 \times 10^{9}M_{\odot}$ and $r_{M87}=16.8 Mpc$. Furthermore, the tidal charge of M87 is taken as $Q_{M87}=9.35 \times 10^{22} C$. 
With the help of the magnetic dipole radiation torque of neutron star, we can calculate the charge parameters using the observational data listed above. In general, the magnetic dipole radiation torque is given by
\begin{equation}
I \frac{d\Omega}{dt} = - \frac{2}{3}\mu^2 \sin^2 \phi \Omega^3,
\end{equation}
in which $\mu$ is the magnetic dipole moment, $\Omega$ is the angular velocity of the star, $I$ is the moment of inertia and $\phi$ is the angle between the rotation and magnetic axis \cite{q}. When we put $\Omega=\frac{2\pi}{P}$ into the Eq.(37) and take $sin\phi=1$, the magnetic dipole moment can be written as
\begin{equation}
\mu=\sqrt{\frac{3IP\dot{P}}{8\pi^{2}}},
\end{equation}
If we combine Eq.(38) and Eq.(16) by considering $I=\frac{Ma}{\Omega}$, the charge parameter can be calculated as
\begin{equation}
e=\frac{1}{a}\sqrt{\frac{3MaP\dot{P}}{8\Omega\pi^{2}}}.
\end{equation}

As Eq.(39) is applicable in geometrical units, we first converted observational data provided in Table I to geometrical units via multiplying the mass, angular velocity and period of each astronomical object by $Gc^{-2}$, $c^{-1}$ and $c$, respectively. This step was then followed by converting the concerned charge outcomes into coulomb by using the conversion factor $G^{-1/2}c^{2}\left( 4\pi
\varepsilon _{0}\right) ^{1/2}$. Note that the gravitational constant, the speed of light and the permittivity of free space are respectively taken as $G=6.67408\times
10^{-11}m^{3}kg^{-1}s^{-2}$, $c=3\times
10^{8}ms^{-1}$ and $\varepsilon _{0}=8.85418\times
10^{-12}C^{2}N^{-1}m^{2}$. In Table II, one can find a list of the numerical values of charges in SI units, as desired.

\begin{table}[H]
\caption{The numerical values of the charges $(e)$ of the relating compact stars.} 
\label{table:nonlin2}
\centering
\begin{tabular}{|l|c|c|r|}
\hline
Compact Stars & $e$  $(C)$ \\ \hline\hline
PSR J 1748-244ad  & $7.249\times 10^{11}$  \\ \hline
PSR B 1937+21  & $4.462\times 10^{11}$  \\ \hline
PSR J 1909-3744  & $1.969\times 10^{11}$ \\ \hline
PSR J 0737-3039 A  & $3.918\times 10^{12}$  \\ \hline
PSR B 1913+12  & $2.581\times 10^{13}$  \\ \hline
\end{tabular}
\end{table}

\begin{figure}[H]
\centering
  \begin{tabular}{@{}cccc@{}}
    \includegraphics[width=.35\textwidth]{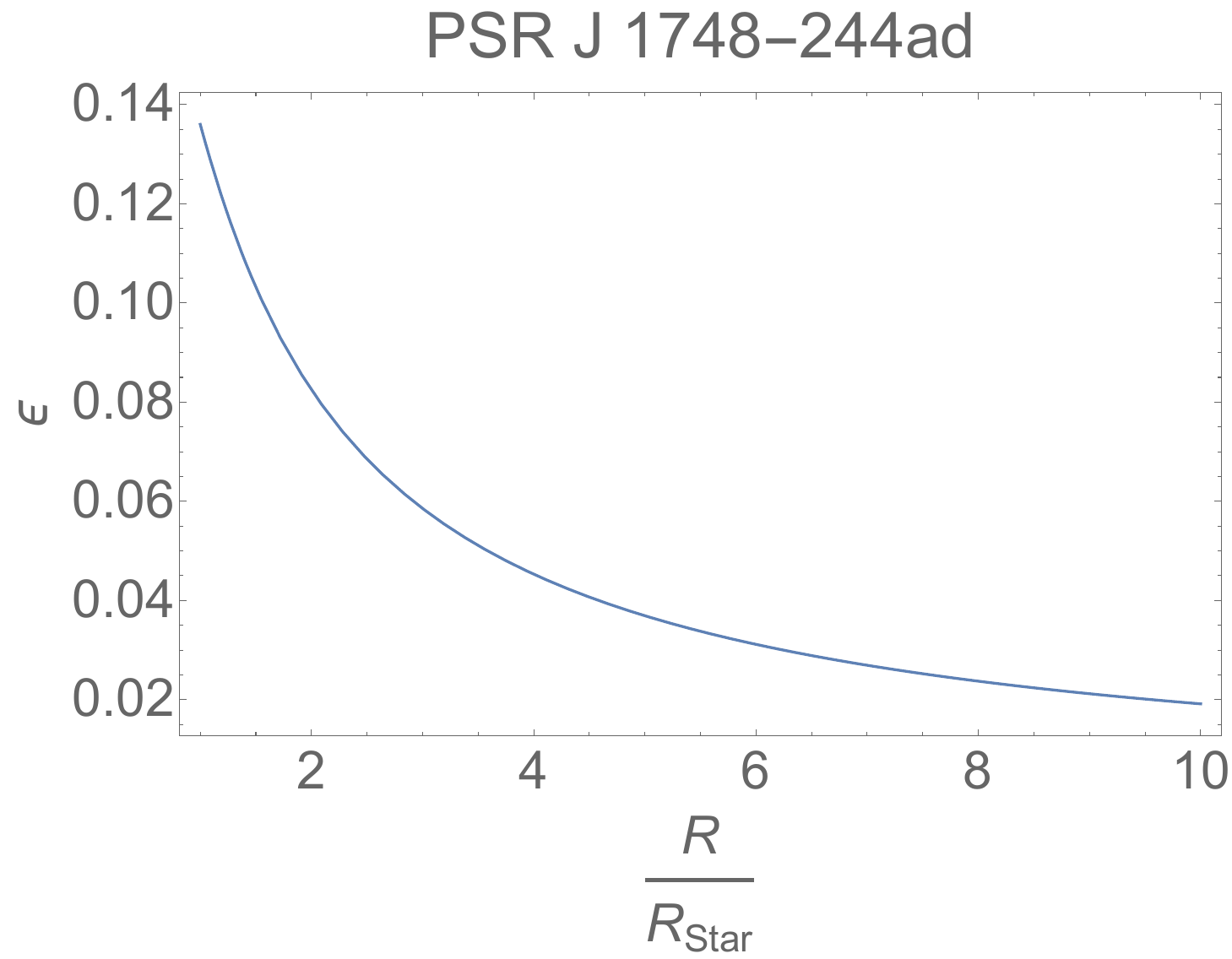} &
    \includegraphics[width=.35\textwidth]{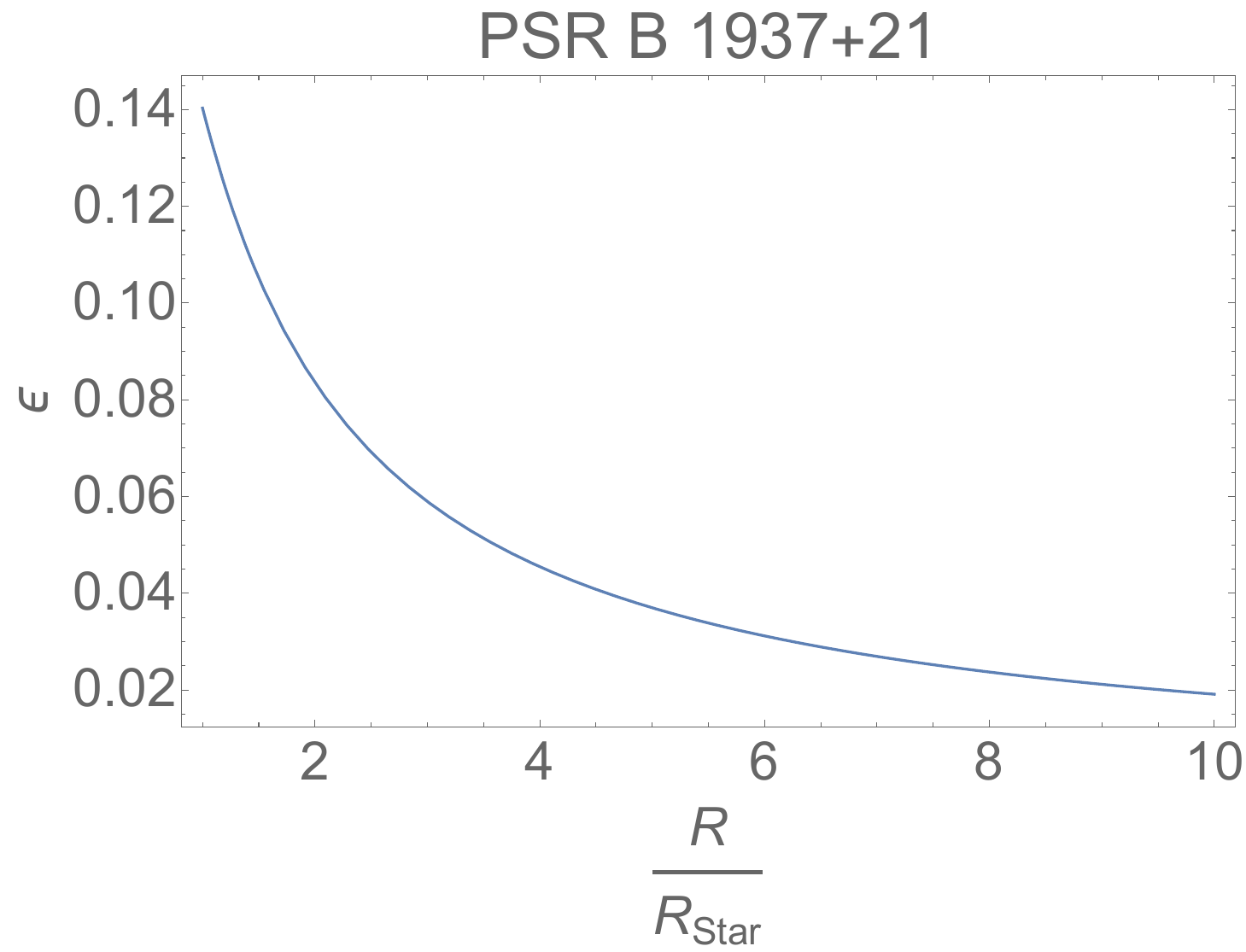} &
  
                                                          \\
   \includegraphics[width=.35\textwidth]{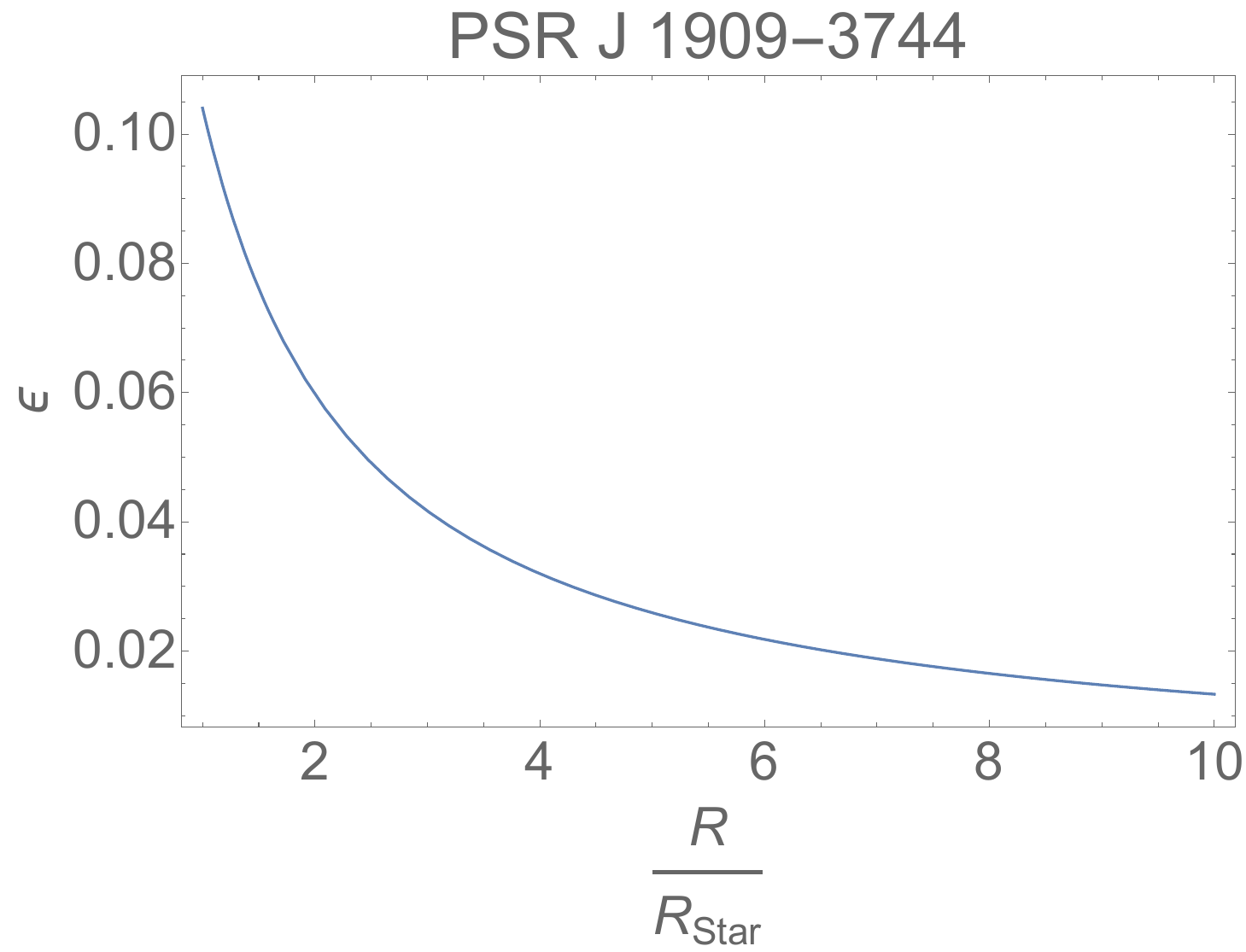} &
    \includegraphics[width=.35\textwidth]{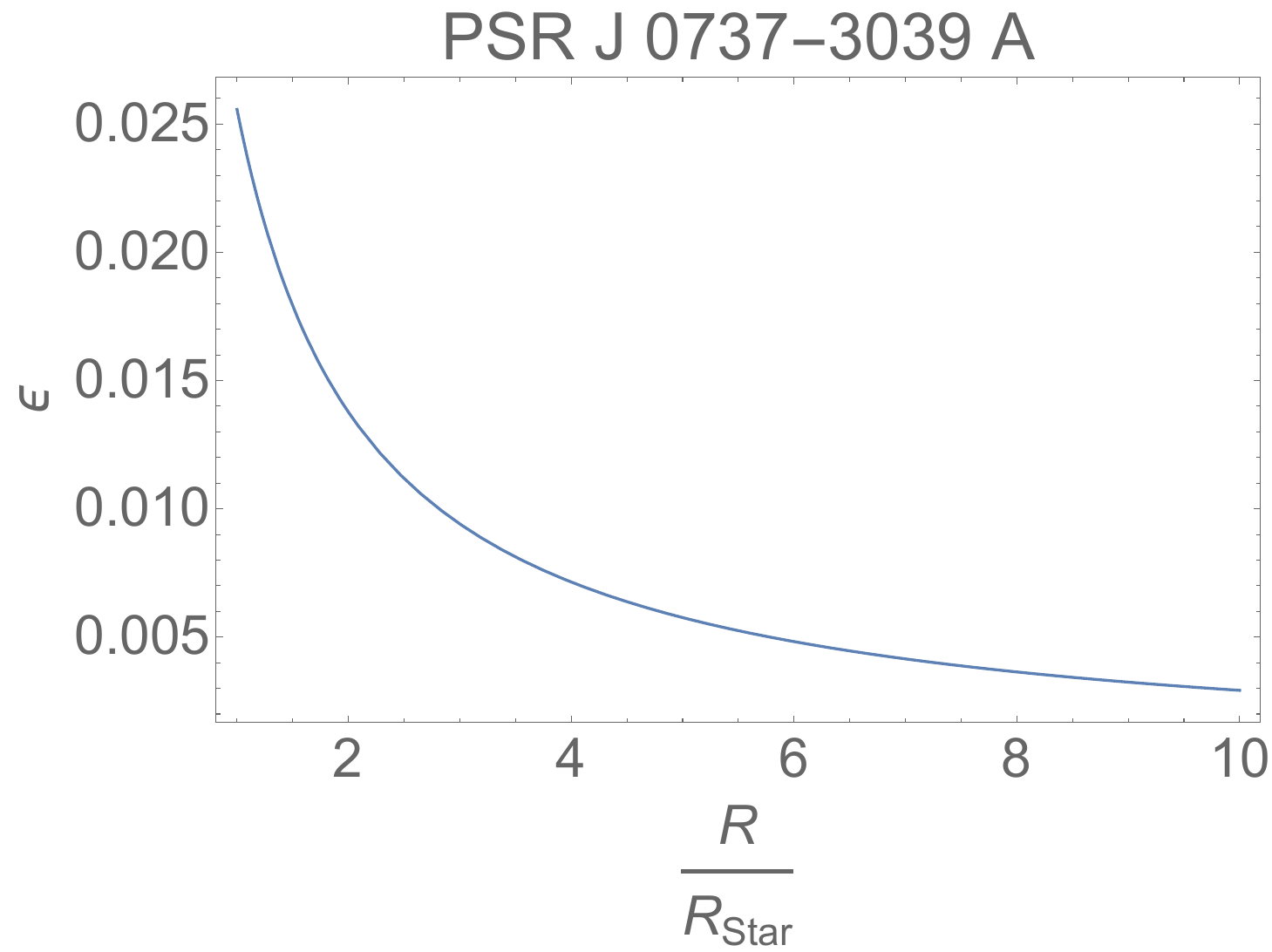} &
                                                             \\
    \multicolumn{2}{c}{\includegraphics[width=.35\textwidth]{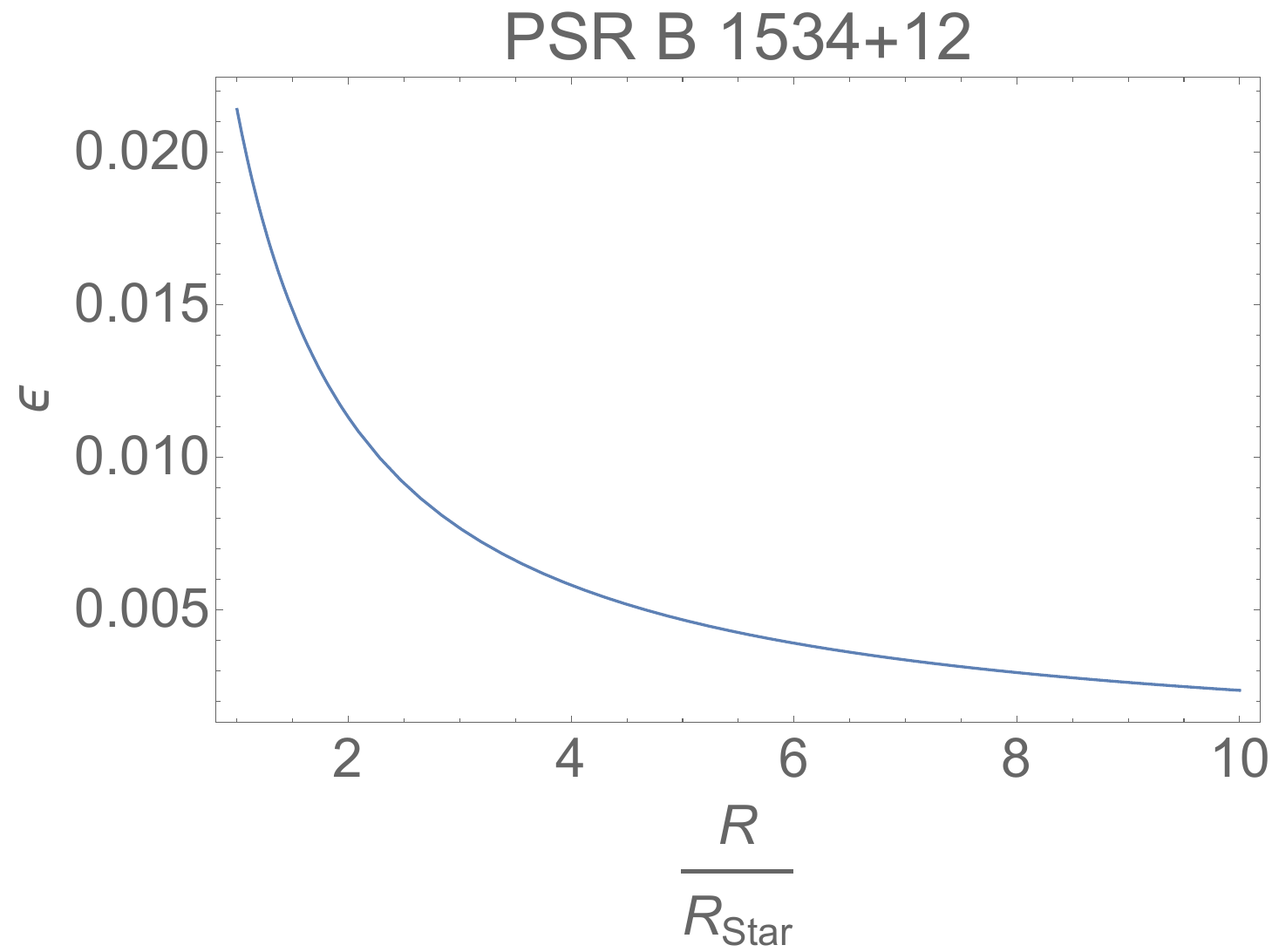}}
  \end{tabular}
  \caption{The effects of two different situations on the bending angle are revealed by taking $R/R_{Star}$, where $R_{Star}$ is the radius of the star in each graph.} \label{fig2}
\end{figure}

The graphical representations of the bending angles for these cases can be viewed in Fig. \ref{fig2}. For the charged compact stars, small angle approximation $(\tan\epsilon\sim\epsilon)$ is used and the graphs are drawn accordingly, whereas for the black holes no limitation has been applied on the angles of concern.
 \\

 Finally, the rotation parameter for black holes is defined as $a=jGM/c^{2}$ with dimensionless rotating parameter having the Kerr bound $j \leq 1$. The aforementioned unit conversions enable us to obtain bending angle of light in SI units as well. 
Graphs in Fig. \ref{fig2} illustrate the variations in the bending angle against a distance for the charged compact stars. The associated graphs are plotted using the estimated numerical values of mass, radius, rotation parameter and charge values for each compact star provided in Table \ref{table:nonlin1} and Table \ref{table:nonlin2}. 

\begin{figure}[H]
\begin{tabular}{cc}
\includegraphics[scale=0.45]{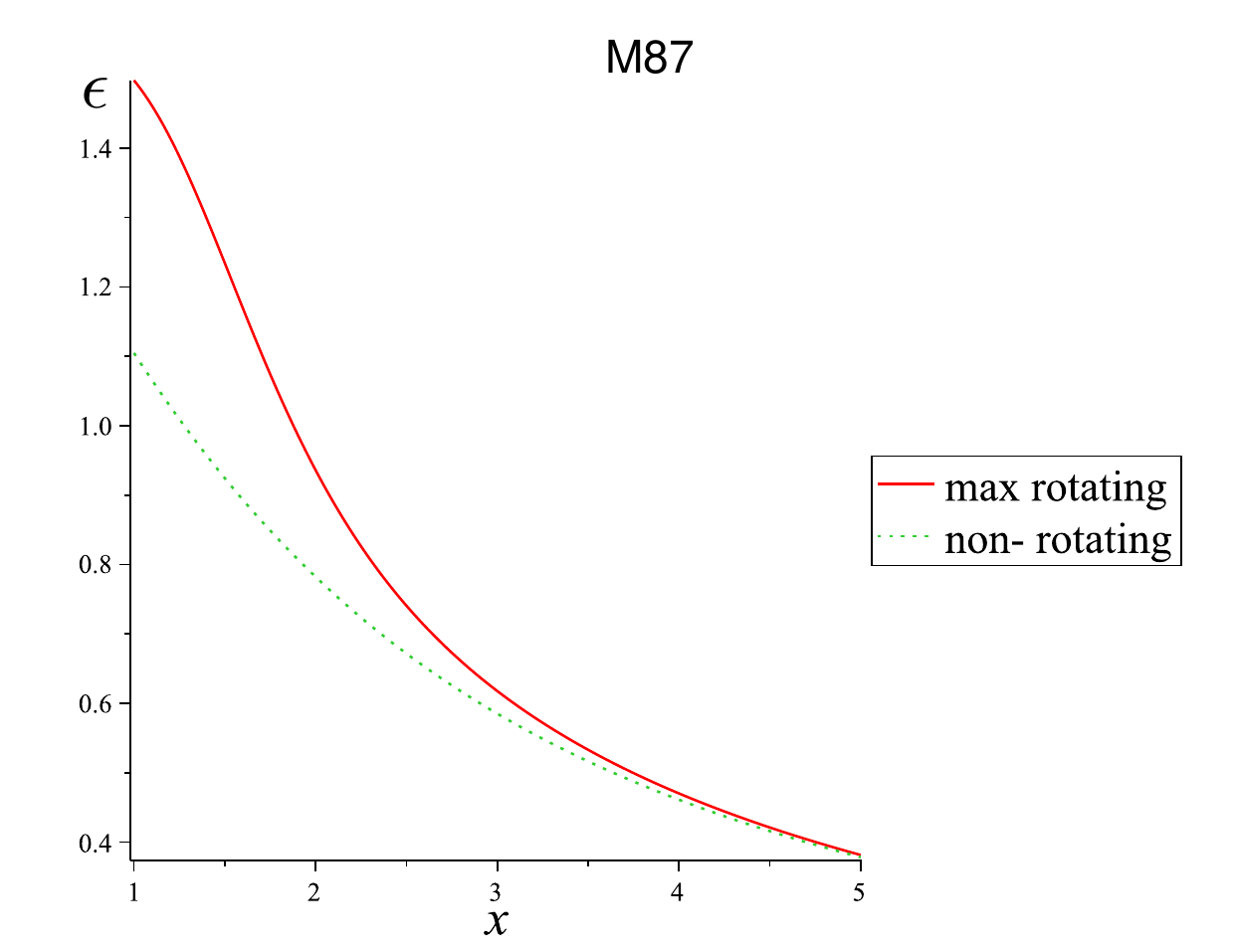}
\includegraphics[scale=0.45]{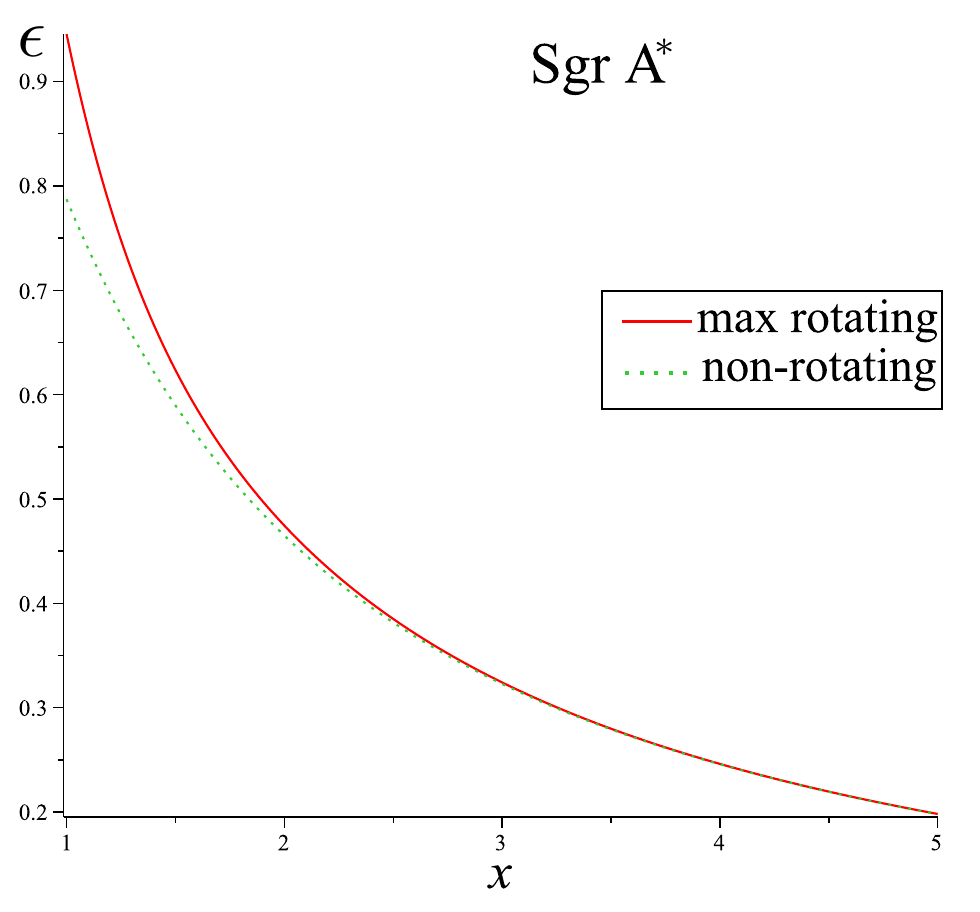}
 \end{tabular}
\centering
 \label{figure}\caption{The bending angles $\epsilon $ versus $x=R/R_{star}$ are plotted for both $M87$ and $Sgr$ $A^{*}$. For $Sgr$ $A^{*}$, the graphs are plotted by assuming $M=4.1\times 10^{6} M_{\odot }$. The Schwarzschild radius is taken as $ 1.27 \times 10^{10} m $ and the charge is considered to be $\approx10^{15}C.$ For $M87$ black hole, the graphs are plotted by assuming that its mass, radius (observable), and tidal charge is $6.5\times 10^{9} M_{\odot }$, $  16.8 Mpc $, and $9.35\times10^{22}C$, respectively. Furthermore, as stated before, $j=1 $ represents the max-rotating case, whereas $j=0 $ illustrates the non-rotating scenario.} \label{fig5}
\end{figure}
\begin{figure}[H]
\centering
  \begin{tabular}{@{}cccc@{}}
    \includegraphics[width=.36\textwidth]{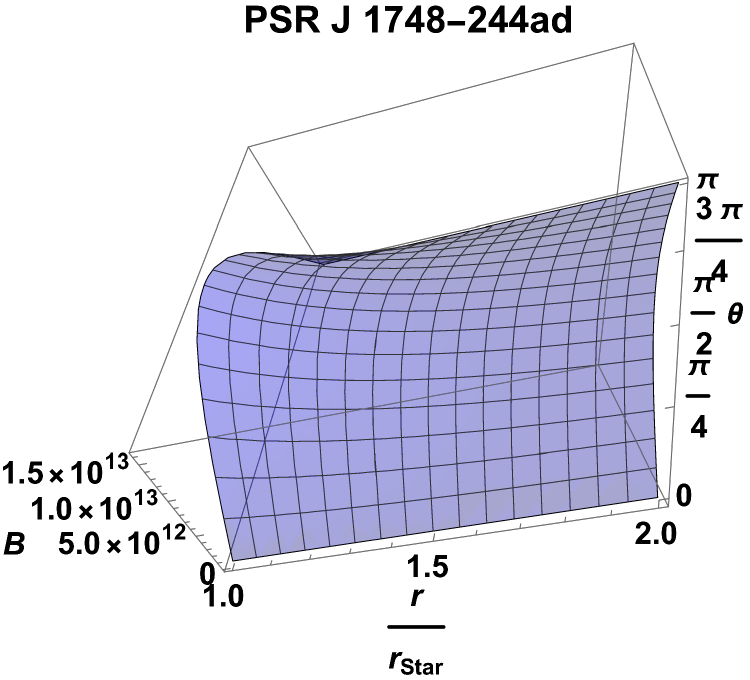} &
    \includegraphics[width=.36\textwidth]{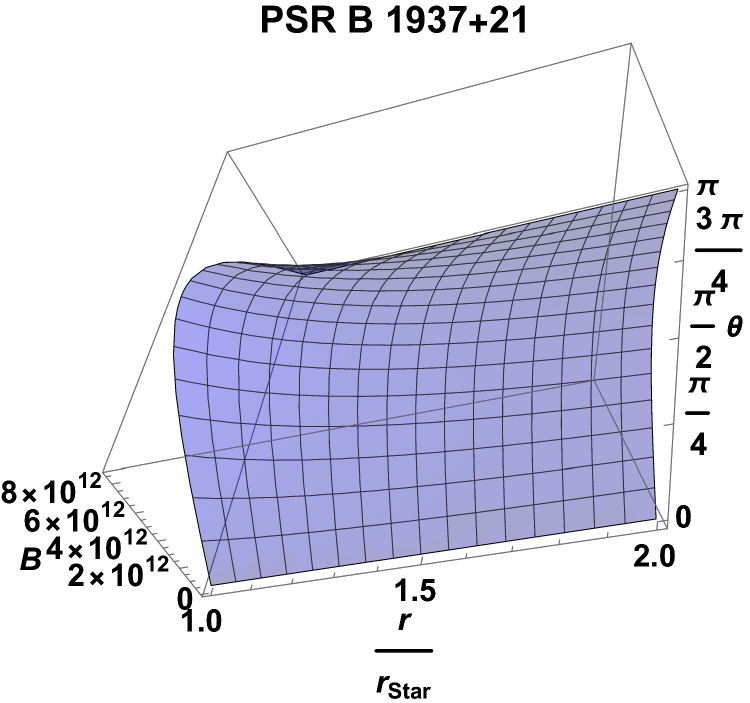} &
                                                         \\
    \includegraphics[width=.36\textwidth]{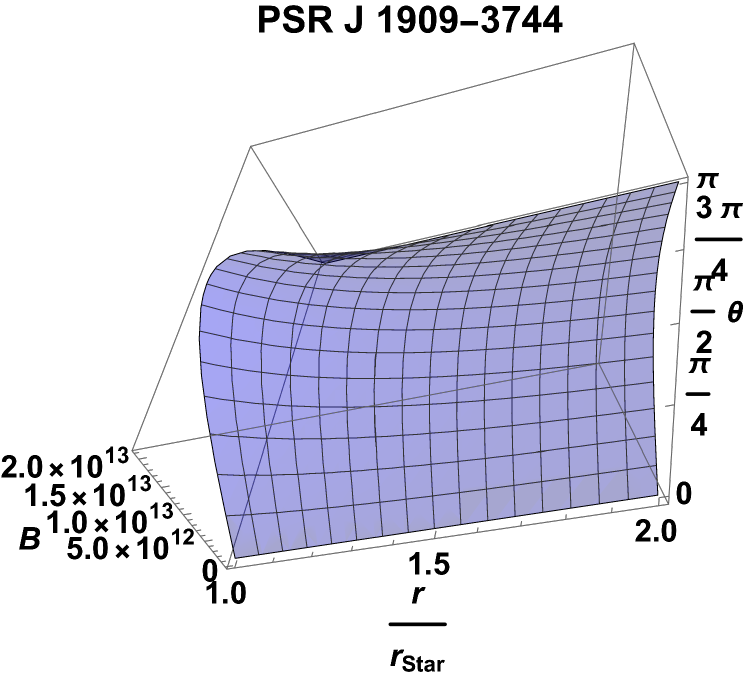} &
    \includegraphics[width=.36\textwidth]{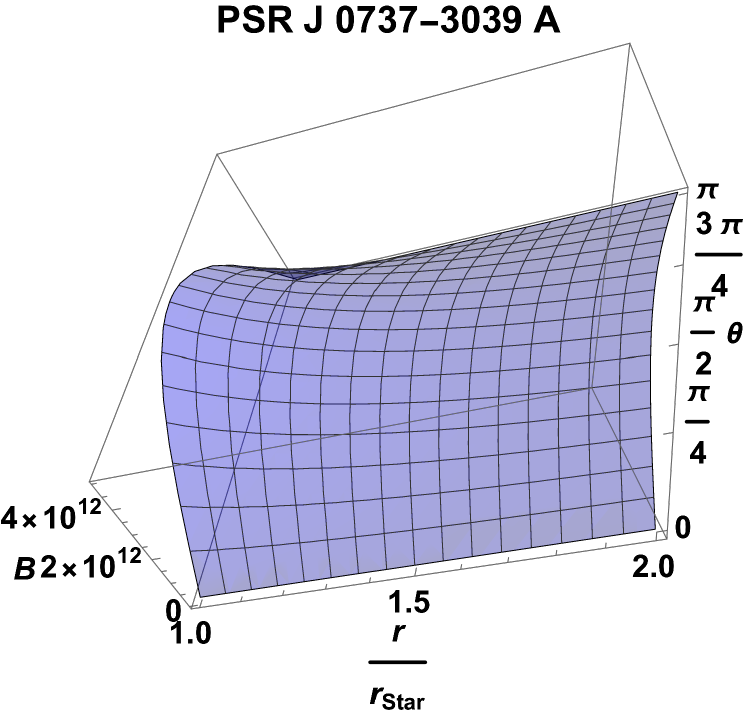} &
                                                            \\
    \multicolumn{2}{c}{\includegraphics[width=.36\textwidth]{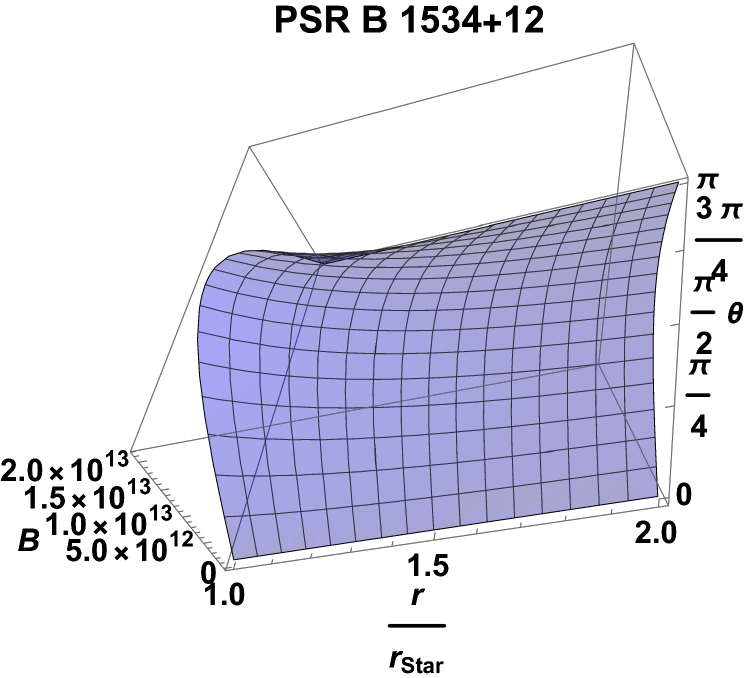}}
  \end{tabular}
  \caption{These figures represent magnetic field strength of stars listed in Table \ref{table:nonlin1} for different $\theta$ values. The analysis is conducted in Gaussian unit system.}  \label{fig6}
\end{figure}

With the intention of inspecting the magnetic field strengths generated due to the rotational effects of the concerned neutron stars in the presence of an KNAdS background, we have used the magnitude of the magnetic field expression having the components of \eqref{11} and \eqref{12}. If one wishes to express the magnetic field in standard units, the required conversion factor is $G^{-1/2}c
\varepsilon _{0} ^{-1/2}$. On the other hand, if Gauss is preferred as the desired unit, an additional factor of $10^4$ is required. For the consequent graphical representations, one can check Figs. \ref{fig6}   

Having obtained the visual representations, it is noteworthy to check whether our results are consistent with the astronomical observations. According to \cite{bha}, the surface magnetic field strengths of neutron stars are expected to range from $10^8$ to $10^{13}$ G. As can be seen in Figs. \ref{fig6}, the maximum value we obtained is $2\times10^{13}$ G, whereas the minimum value reads $4\times10^{12}$ G. Therefore, our results are consistent with both the theoretical and observational expectations \cite{bha,teusolky,pengy}. We have used the Gaussian unit system for the magnetic field evaluations only, as we wanted to compare our magnetic field range with the aforementioned limits in literature. Our results also imply that as the neutron star rotates faster, magnetic field strength increases accordingly.  
\section{Results and Discussions} \label{sect6}

In this study, the deflection of light in the spacetime of a rotating electrically charged black hole (Kerr-Newman) has been studied including the contribution from the $\Lambda$. Having provided a brief summary on the general formula applicable to rotating spacetimes with curved background, we have calculated the amount of one-sided deflection for the KNAdS spacetime. The method we have used was obtained via the application of RIM to spacetimes with rotation. \\

In addition to the general bending angle expression, more specific situations have also been considered by concentrating on the effect of mass, electric charge and rotational parameters of a set of astronomical objects on the path that light follows. For inspecting observational contributions, the aforementioned scenarios are checked for numerous observational data belonging to seven different compact neutron stars ($PSR J 1748-2446ad,$ $PSR B 1937+21$, $PSR J 1909-3744$, $PSR J 0737-3039 A$ and $PSR B 1534+12$) and two black holes (Sgr $A^{*}$ and M87). Our computations have shown that the contributions of mass and rotation are prominent, whereas the electric charge seems to have a relatively weak effect on the deflection of light. Furthermore, as rotation parameter increases, the deflection angle also increases.  
\\

When the rotation and charge parameters are chosen as zero, our bending angle matches the one obtained by Ishak and Rindler in Schwarzschild-de Sitter geometry. 
We can also comment that $\Lambda$ and the rotational term seem to have a coupled effect on the overall bending, whereas distinct contributions arise due to the charge parameter. Further comments can be made on the rotation parameter and its relevance to the bending. As the astronomical object at the center rotates faster, an increase in the deflection angle is generated. The effect is more apparent in Sgr $A^{*}$ and M87 than those evaluated for the neutron stars. Therefore, one could conclude rotation of black holes influence the light ray paths to a greater extent. On the other hand, $\Lambda$ also plays an important role in final expression attained for the closest approach distance which also seems to be linked to the rotation in presence.

Modified gravity is believed to be a good candidate to explain unknown components such as dark energy and dark matter in the universe. Rastall gravity \cite{Rastall}, one of the modified gravity theories, gives interesting results in which the universe is assumed to be consisting of interacting/non-interacting dark energy and dark matter \cite{izson1}. Recently, KNAdS black hole surrounded by perfect fluid matter in the Rastall gravity has been derived \cite{izson2}. Therefore, it would be intriguing to extend our study for more observables to reveal the effect of the Rastall gravity and hence the dark matter/energy on the gravitational lensing phenomena. Besides, another work that we would like to consider in the future is the problem of immersing of KNAdS black hole into an external magnetic field. As is well known \cite{Karas:2014paa}, for rotating black holes, magnetic fields never penetrate the horizon, which is specifically at odds with a famous magnetosphere solution found in 1974 by Robert M. Wald \cite{wald74} and it is rejected by the black hole, just as a superconductor rejects a magnetic field: Missner effect \cite{misn}. However, this would require analytical solution of test Maxwell's equations in the KNAdS background. However, this study will require a difficult and separate study that will require analytical solution of the test Maxwell equations in the KNADS background. In particular, it would be interesting to be able to reveal the effect of $\Lambda$ on the Missner effect, if there is any. Furthermore, expanding our evaluations to higher orders including the second- order solutions to the null geodesic equation also sounds intriguing. By this way, we would be able to compare the GL values of this paper with the ones to be obtained following the perturbative methods mentioned in \cite{new7} and \cite{new5}. Those issues are going to be considered in our future work agenda. Meanwhile, with the captures of M87 \cite{EventHorizonTelescope:2019dse} and Sgr $A^{*}$ \cite{EventHorizonTelescope:2022xnr}, it would not be an exaggeration to claim that these two black holes will be scrutinized closely for a longer period of time. In this context, we expect that the gravitational lensing phenomena reported in this paper will help the observers to explain the future findings. Finally, the existence of quark matter at the cores of the massive neutron stars has been suggested by holographic models \cite{BitaghsirFadafan:2020otb}. This intriguing result supports the new light-quark dark matter hypothesis \cite{Bashkanov:2020dfb}. As a result, we shall keep contributing to the gravitational lensing research of compact objects since they probe the amount and nature of the dark matter.

\section*{Acknowledgements}
We are thankful to the Editor and anonymous Referee for their constructive suggestions and comments. \.{I}. Sakall{\i} gratefully acknowledge the contributions of TÜBİTAK and SCOAP3. He also thank to Prof. Behnam Pourhassan for fruitful dicussions.

\end{document}